\documentclass[aps,prl,twocolumn,groupedaddress]{revtex4-1}
\usepackage{amsmath}
\usepackage{amssymb}
\usepackage{graphicx}
\usepackage{dcolumn}
\usepackage{bm}

\DeclareMathOperator{\sinc}{sinc}

\begin{document}

\title{Metasurfaces with bound states in the continuum enabled by eliminating first Fourier harmonic component in lattice parameters}

\author{Sun-Goo Lee}
\email{sungooleee@gmail.com}
\author{Seong-Han Kim}
\author{Chul-Sik Kee}
\email{cskee@gist.ac.kr}
\affiliation{Integrated Optics Laboratory, Advanced Photonics Research Institute, Gwangju Institute of Science and Technology, Gwangju 61005, South Korea}
\date{\today}

\begin{abstract}
Conventional photonic lattices, such as metamaterials and photonic crystals, exhibit various interesting physical properties that are attributed to periodic modulations in lattice parameters. In this study, we introduce novel types of photonic lattices, namely Fourier-component-engineered metasurfaces, that do not possess the first Fourier harmonic component in the lattice parameters. We demonstrate that these metasurfaces support the continuous high-$Q$ bound states near second stop bands. The concept of engineering Fourier harmonic components in periodic modulations provides a new method to manipulate electromagnetic waves in artificial periodic structures.
\end{abstract}


\maketitle

Subwavelength photonic lattices with thin-film geometries, such as metasurfaces \cite{Kildishev2013,NYu2014,SSun2019} and photonic crystal slabs \cite{Joannopoulos1995,Johnson1999}, have attracted significant attention in recent years owing to their substantial ability to manipulate electromagnetic waves. Unlike the usual thin homogeneous dielectric layers governed by Fresnel equations and Snell's law \cite{Born2002}, photonic lattices  can capture incident light by resonance through lateral Bloch modes and reemit it with predesigned electromagnetic responses \cite{YHKo2018}. By appropriately designing individual constituents in lattices, several interesting physical effects and useful applications, which cannot be achieved with conventional dielectric materials, can be realized in an extremely compact format, even as a single-layer film \cite{Arbabi2015,AMHWong2018,Niraula2015}.

Recently, extensive studies were performed on bound states in the continuum (BICs) with exceptionally high radiative $Q$ factors in one-dimensional (1D) and 2D planar photonic lattices. In principle, BICs with infinite $Q$ factors can be found as unusual electromagnetic eigenstates that remain well localized in open photonic systems, even though they can coexist with the continuous spectrum of outgoing waves \cite{Marinica2008,Plotnik2011,Hsu2016,Koshelev2019,SGLee2020-1}. BICs are associated with various fascinating physical phenomena, such as the sharp Fano resonances \cite{Koshelev2018}, enhanced nonlinear effects \cite{Koshelev2020}, and topological nature \cite{BZhen2014,Doeleman2018}. Different types of BICs have been studied in versatile planar photonic lattices \cite{Azzam2018,Yang2014,Mermet-Lyaudoz2019,JJin2019,XYin2020}. However, the high-$Q$ BICs introduced in the literature thus far are very sensitive to the wavevector of Bloch modes in the lattices. Hence, sharp Fano resonances and enhanced nonlinear effects due to the high-$Q$ Bloch modes can be obtained at a discrete specific incident angle. Small variations in the incident angle significantly reduce the resonance $Q$ factor in spectral responses.

In this Letter, we introduce the concept of Fourier-component-engineered (FCE) metasurfaces that do not possess the first Fourier harmonic component in the periodically modulated lattice parameters and demonstrate that the metasurfaces can support the continuum of high-$Q$ bound states in a wide range of wavevectors instead of a specific discrete wavevector. Guided modes in 1D or 2D FCE metasurfaces exhibit noticeably increased radiative $Q$ factors, as out-of-plane radiation occurs due to the first Fourier harmonic component of periodic modulations. Two types of FCE metasurfaces that utilize the spatially engineered dielectric function and thickness profile, respectively, are presented and analyzed.

\begin{figure}[t]
\centering
\includegraphics[width=8.0cm]{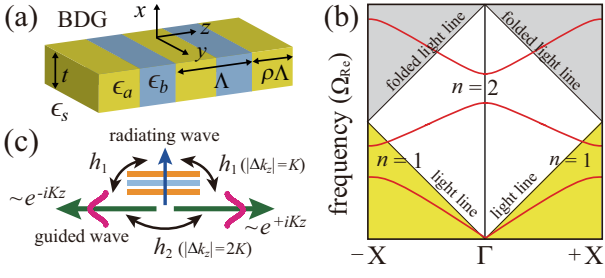}
\caption {\label{fig1} (a) Schematic of conventional 1D BDG that supports leaky guided modes. (b) Photonic band gaps. (c) Illustration of coupling processes that induces a second band gap.}
\end{figure}

Figure~\ref{fig1}(a) illustrates the simplest representative 1D photonic lattice, i.e., binary diffraction grating (BDG), consisting of high ($\epsilon_a$) and low ($\epsilon_b$) dielectric constant materials. The thickness of the grating is $t$, and the width of high dielectric constant part is $\rho \Lambda$, where $\Lambda$ is the period of the grating. The BDG layer acts as a waveguide, as well as a diffracting element, because its average dielectric constant $\epsilon_{avg}=\rho \epsilon_{a}+(1-\rho) \epsilon_{b}$ is larger than that of surrounding medium ($\epsilon_{s}$). The periodic modulations in the dielectric constant generate photonic band gaps at the Bragg condition, as shown in Fig.~\ref{fig1}(b). The first stop band ($n=1$) in the yellow region is not associated with the leaky-wave effects, as it is protected by total internal reflection. Near the second stop band ($n=2$) in the white region, leaky-wave effects occur, because Bloch modes are described by complex eigenfrequencies $\Omega = \Omega_{\mathrm{Re}} + i\Omega_{\mathrm{Im}}$. In this study, we focus on the second band gap of the fundamental $\mathrm{TE}_{0}$ mode, because high-$Q$ BICs are associated with the second stop bands in numerous cases. As conceptually illustrated in Fig.~\ref{fig1}(c), the second leaky stop band is formed primarily by the direct coupling $h_2$ with $|\Delta k_z| = 2K$ between two counterpropagating waves $\sim\exp (\pm i Kz)$, where $K=2 \pi/\Lambda$ is the magnitude of the grating vector, and secondarily by the radiative coupling $h_1$ with $|\Delta k_z| = K$ between the guided and radiating waves \cite{YDing2007}.

We first show that the coupling processes $h_1$ and $h_2$ are associated with the first and second Fourier harmonic components, respectively, and the out-of-plane radiation loss is caused by the first Fourier harmonic. The eigenfrequencies of the Bloch modes in the BDGs can be obtained by solving the 1D wave equation given by \cite{Yariv1984}:
\begin{equation}\label{wave-equation}
\left (\frac{\partial^{2}}{\partial x^{2}}  + \frac{\partial^{2}}{\partial z^{2}} \right ) E_{y}(x,z) + \epsilon (x,z) k_{0}^2 E_{y}(x,z)= 0,
\end{equation} 	 	
where $k_{0}$ denotes the wave number in free space. Equation~(\ref{wave-equation}) can be solved by expanding the periodic dielectric function $\epsilon (x,z)$ in a Fourier series and the electric field $E_{y}$ as a Bloch form \cite{Inoue2004}. For the BDG with inversion symmetry, the dielectric function can be expanded in an even cosine function series $\epsilon(z)=\sum_{0}^{\infty} \epsilon_{n}\cos(nKz)$, where the Fourier coefficients are given by $ \epsilon_{0}=\epsilon_{avg}$ and $ \epsilon_{n\geq1}=(2\Delta \epsilon / n\pi) \sin (n\pi \rho)$. For a clear insight into the formation of the second stop band, we use a simple semi-analytical approach, in which only the zeroth, first, and second Fourier harmonics are retained for the dielectric function, and the spatial electric field distribution is approximated as $E_{y}(x,z)=[A\exp(+iK z)+B\exp(-iK z)]\varphi(x)+E_{rad}$, where $A$ and $B$ are slowly varying envelopes of the two counter-propagating waves, $\varphi(x)$ characterizes the mode profile of the unmodulated waveguide, and $E_{rad}$ represents the radiating diffracted wave \cite{Kazarinov1985}. Solving the wave equation with the approximated dielectric function and field distributions, the dispersion relations near the second stop band can be written as
 \begin{equation}\label{eq2}
\Omega(k_{z})=\Omega_{0} - \left ( ih_{1} \pm \sqrt{k_{z}^2+(h_{2}+ih_{1})^2} \right )/(Kh_{0}),
\end{equation} 	 	
where $\Omega_{0}$ is the Bragg frequency under the vanishing index modulation ($\Delta \epsilon =0$), and the coupling coefficients are given by
\begin{equation}\label{eq3}
h_{0} = \displaystyle \Omega \int_{-\infty}^{\infty} {\epsilon_{0}}(x) \varphi(x)\varphi^*(x) dx,
\end{equation}
\begin{eqnarray}\label{eq4}
h_{1} = i \displaystyle \frac{ K^3\Omega^4 \epsilon_{1}^2}{8} \int_{-t}^{0}  \int_{-t}^{0} G(x,x^{\prime})\varphi(x')\varphi^*(x) dx' dx, 
\end{eqnarray}
\begin{eqnarray}\label{eq5}
h_{2} = \displaystyle \frac{ K\Omega^2 \epsilon_{2}} {4} \int_{-t}^{0}  \varphi(x)\varphi^*(x) dx,
\end{eqnarray}
where $G(x,x')$ denotes the Green's function for the diffracted field \cite{SGLee2019-1,YDing2007,Rosenblatt1997}. The dispersion relations in Eq.~(\ref{eq2}) can be obtained by calculating the coupling coefficients in Eqs.~(\ref{eq3})--(\ref{eq5}). Details of the mathematical process used to obtain Eq. (\ref{eq2}) is provided in the Supplemental Material \cite{Supplemental}.

Equation~(\ref{eq2}) indicates that the leaky stop band with two band edges $\Omega^a=\Omega_{0}+h_{2}/(\mathrm{K}h_{0})$ and $\Omega^s=\Omega_{0}-(h_{2}+i2h_{1})/(\mathrm{K}h_{0})$ opens at $k_{z} = 0$. For the conventional BDG with inversion symmetry, the coupling coefficient $h_1$ is generally a complex value, whereas $h_0$ and $h_2$ are real values. Except for the symmetry-protected BIC with purely real frequency $\Omega^a$, Bloch modes near the second stop band generally suffer radiation loss, because they have complex frequencies, as indicated in Eq.~(\ref{eq2}). Because the second stop band opens at the second-order $\Gamma$ point ($k_z=K$ in the extended Brillouin zone) with the Bragg condition $k_z = q(\pi/p)$, where $p$ is the period of the dielectric constant modulation, and $q$ represents the Bragg order, it is reasonable to interpret that the coupling coefficients $h_1$ and $h_2$ represent the second-order Bragg effect due to the first Fourier harmonic, $\epsilon_{1}\cos(Kz)$, and the first-order Bragg effect due to the second Fourier harmonic, $\epsilon_{2}\cos(2Kz)$, respectively. Without $h_1$, in the vicinity of the $\Gamma$ point, guided modes become BICs irrespective of $k_z$ with the real eigenfrequencies given by $\Omega(k_{z})=\Omega_{0} \pm \sqrt{k_{z}^2+h_{2}^2} /(Kh_{0})$. Inspired by the analytical dispersion relations, we introduce and analyze FCE metasurfaces without the first Fourier harmonic component through rigorous finite element method (FEM) simulations. Since the coupling coefficient $h_1$ is due to $\epsilon_{1}\cos(Kz)$, we make the $h_1$ zero by eliminating the first Fourier harmonic component from the original step-like dielectric functions.

\begin{figure}[]
\centering
\includegraphics[width=8.0 cm]{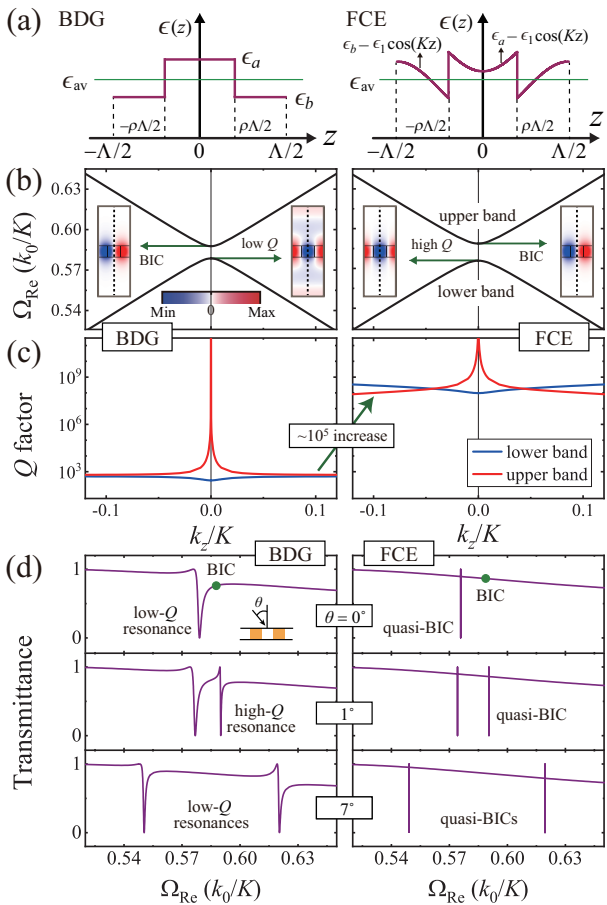}
\caption{\label{fig2} Comparison between the conventional BDG and FCE metasurface. (a) Dielectric functions with respect to $z$. (b) FEM-simulated dispersion relations. Insets with blue and red colors illustrate the spatial electric field ($E_{y}$) distributions of band edge modes at the $y=0$ plane. The vertical dotted lines represent mirror planes in computational cells. (c) Calculated radiative $Q$ factors of the upper and lower bands. (d) Evolution of transmission spectra vs. incident angle $\theta$. In FEM simulations, we use the structural parameters $\epsilon_{avg}=4.00$, $\Delta\epsilon=1.00$, $\epsilon_{s}=1.00$, $t=0.50~\Lambda$, and $\rho = 0.40$.}
\end{figure}

In Fig.~\ref{fig2}, we compare the key properties of the conventional BDG and those of the corresponding FCE metasurface. As shown in Fig.~\ref{fig2}(a), the FCE metasurface has complex dielectric functions $\epsilon_a - \epsilon_{1} \cos (Kz)$ and $\epsilon_b - \epsilon_{1} \cos (Kz)$, when $|z| < \rho\Lambda /2$ and $|z| \geq \rho\Lambda /2$, respectively, whereas the conventional BDG has simple step-like dielectric functions with $\epsilon_a$ and $\epsilon_b$. The simulated dispersion relations illustrated in Fig.~\ref{fig2}(b) show that the second bandgap opens at $k_z=0$ for both the conventional BDG and the FCE metasurface. Dispersion curves for the BDG and the FCE metasurface seem similar, and the spatial electric field ($E_y$) distributions in the insets show that symmetry-protected BICs with asymmetric field distributions appear at upper band edges in both the BDG and the FCE metasurface. However, a noticeable difference between the conventional BDG and the FCE metasurface can be observed from the symmetric spatial electric field distributions of the lower edge mode. While the lower edge mode in the BDG is radiative out of the lattice, that in the FCE metasurface is appropriately localized in the lattice, even though it is not protected by the symmetry mismatch. The effect of the first Fourier harmonic component is likewise clearly seen by investigating the radiative $Q$ factors as a function of $k_z$ plotted in Fig.~\ref{fig2}(c). In the BDG, the BIC in the upper band exhibits a $Q$ factor that is larger than $10^{15}$ at the $\Gamma$ point, whereas the $Q$ values decrease abruptly and approach the value of the $Q$ factor of leaky modes ($\sim10^{3}$) in the lower band, as $k_z$ moves away from the $\Gamma$ point. Similarly, in the FCE metasurface, the BIC also exhibits a $Q$ factor that is larger than $10^{15}$, and the $Q$ values decrease as $k_z$ moves away from the $\Gamma$ point. However, the Bloch modes in both the upper and lower band branches have high $Q$ values ($\sim10^{8}$) in the computational range of $|k_z|\leq0.12~K$. The $Q$ factors in the FCE metasurface are approximately $10^5$ times larger than those in the conventional BDG with the same lattice parameters except for the profile of the dielectric function.

The symmetry-protected BICs at the $\Gamma$ point are perfectly embedded eigenvalues with an infinite $Q$ factor and vanishing resonance line width, because they are perfectly decoupled from external waves. In diverse practical applications, quasi-BICs with finite high-$Q$ values and narrow spectral responses via resonant coupling with external waves are favorable. Figure~\ref{fig2}(d) illustrates the transmission spectra through the BDG and FCE metasurface for three different values of incident angles $\theta =0^\circ$, $1^\circ$, and $7^\circ$. At normal incidence with $\theta =0^\circ$, the BDG structure exhibits only the low-$Q$ resonance by the lower band edge mode. The embedded BIC in the upper band edge mode, a green solid circle in the transmittance curve, does not generate the resonance effect. When $\theta =1^\circ$, the simulated transmittance curve through the conventional BDG exhibits low-$Q$ resonance by the lower band mode and high-$Q$ resonance in the upper band mode. As $\theta$ increases further, the resonance line width in the upper band mode in the BDG increases rapidly and becomes close to that of the lower band mode. Two low-$Q$ resonances by the upper and lower band modes are observed in the spectral response when $\theta =7^\circ$. Similarly, the transmittance curve through FCE metasurface does not show the embedded BIC in the upper band when $\theta =0^\circ$. However, the resonance line width in the lower band mode in the metasurface is extremely narrow, irrespective of the incident angle $\theta$, unlike the case of conventional BDG. As $\theta > 0$ increases, two quasi-BICs by the lower and upper band modes are observed in the transmission curves, and the resonance line widths remain narrow when $\theta =7^\circ$. Numerous efforts have been devoted to realizing high-$Q$ resonant modes in photonic crystal structures. However, high-$Q$ modes reported in previous studies were obtained at specific discrete wavevectors \cite{Srinivasan2002}. Because the proposed FCE metasurface can support continuous quasi-BICs in the vicinity of the second stop bands, they could find useful applications to overcome the discrete nature of the high-$Q$ resonant mode in conventional photonic lattices. As an example of possible applications of high-$Q$ bound states in the FCE metasurface, a discussion on the Dirac cone dispersion is presented in the Supplemental Material \cite{Supplemental}.

\begin{figure}[t]
\centering
\includegraphics[width=8.5cm]{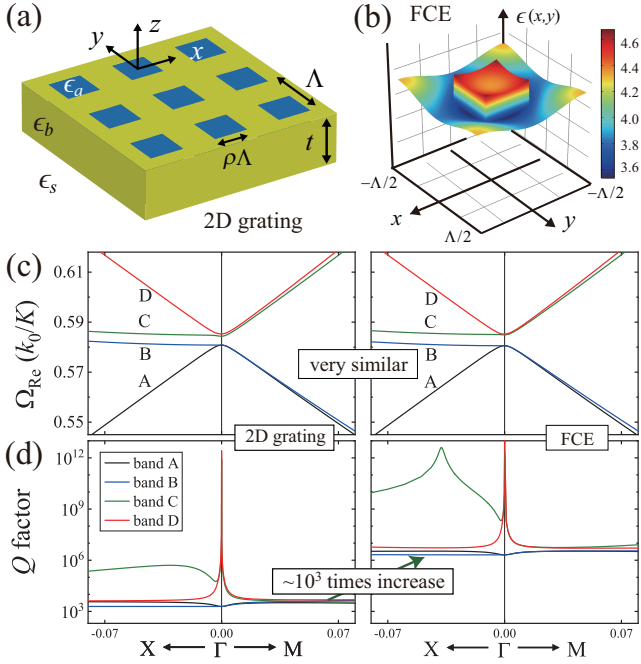}
\caption{\label{fig3} (a) Schematic of 2D grating. (b) Profile of dielectric constant of FCE metasurface corresponding to 2D grating. Simulated dispersion relations (c) and radiative $Q$ factors (d) in conventional 2D grating and FCE metasurface. In the simulations, we used lattice parameters $\Delta \epsilon=\epsilon_a-\epsilon_b=1.00$, $\rho = 0.40$, $\epsilon_{avg} =4.00$, $t=0.50~\Lambda$, and $\epsilon_{s}=1.00$. }
\end{figure}

We now show that the radiative $Q$ factors in the 2D lattices can also be increased by eliminating the appropriate Fourier components in the original dielectric function. As illustrated in Fig.~\ref{fig3}(a), we consider a simple 2D lattice composed of square arrays of square-shaped high dielectric constant ($\epsilon_a$) materials in the background medium with a low dielectric constant ($\epsilon_b$). The dielectric function can be expanded as
\begin{equation}\label{Fourier}
\epsilon(x,y) = \epsilon_{avg} + \sum_{(m,n) \neq (0,0)}^{\infty} \gamma_{m,n} e^{i(mKx+nKy)},
\end{equation} 	 	
where the Fourier coefficients are given by $\epsilon_{avg} = \rho^2 \epsilon_{a}+ (1 - \rho^2)\epsilon_b$ and $\gamma_{m,n} =\rho^2 \Delta \epsilon \times \sinc(m\pi \rho)\times \sinc(n\pi \rho)$. In the 1D case, the first Fourier component with a spatial period of $\Lambda$ was removed. As an extension of the 1D study to the 2D grating structure, we removed the two Fourier harmonic components, $2\gamma_{1,0} \cos (Kx)$ and $2\gamma_{0,1} \cos (Ky)$ with a spatial period of $\Lambda$, from the original dielectric function, as illustrated in Fig.~\ref{fig3}(b). Figure~\ref{fig3}(c) shows that the second band gaps open at the $\Gamma$ point, and the dispersion curves for the conventional 2D grating and FCE metasurface are very similar. There are four bands, A, B, C, and D, in the vicinity of the $\Gamma$ point \cite{Peng2012}. Spatial electric field distributions and radiative properties of the four band edge modes are presented in the Supplemental Material \cite{Supplemental}. The existence of the BICs can be observed from the radiative $Q$ factors plotted in Fig.~\ref{fig3}(d). In both the conventional 2D grating and FCE lattice, the symmetry-protected BICs in bands C and D exhibit $Q$ factors larger than $10^{12}$ at the $\Gamma$ point; however, the $Q$ values decrease noticeably as $k_z$ moves away from the $\Gamma$ point. Nevertheless, the simulated radiative $Q$ factors in the FCE metasurface are approximately $2\times10^3$ times larger than those in the conventional 2D grating.  

\begin{figure}[b]
\centering
\includegraphics[width=8.5cm]{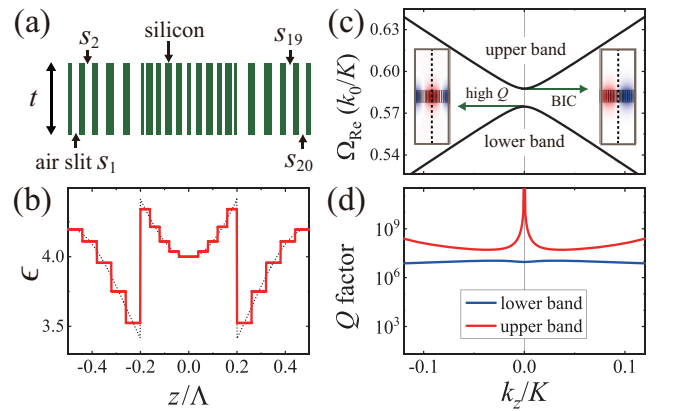}
\caption{\label{fig4} (a) Schematic of a 1D silicon-air metasurface with $t=0.50~\Lambda$. (b) Effective dielectric constant of the silicon-air metasurface in the computational unit cell. The dotted line represents the dielectric constant of the FCE metasurface. Simulated dispersion relations (c) and radiative $Q$ factors (d) near the second stop band of the silicon-air metasurface.}
\end{figure}

Since the dielectric functions of the FCE metasurfaces shown in Fig.~\ref{fig2}(a) and Fig.~\ref{fig3}(b) have some fine details, the practical implementation of the FCE metasurface with the engineered spatial dielectric function is challenging. Currently, the required dielectric constant profile can be implemented in long wavelength regions, such as microwaves, where diverse metamaterials consisting of deep subwavelength size components are feasible \cite{Schurig2006,HFMa2010}. As the nanofabrication technology continues to improve, it could be possible to realize the FCE metasurface operating at optical wavelengths in the future. To visualize this, as shown in Fig.~\ref{fig4}(a), we consider a 1D silicon-air metasurface consisting of air slits, $S_{j}$, in a silicon slab ($\epsilon_{\mathrm{Si}}=12.11$ at $\lambda=1.55~\mu m$). Based on the effective medium theory \cite{Haggans1993}, twenty air slits are appropriately located in the unit cell of size $\Lambda$, such that the effective dielectric function of the silicon-air metasurface mimics that of the FCE metasurface, as illustrated in Fig.~\ref{fig4}(b). Even though the effective dielectric function of the silicon-air metasurface varies in discrete steps, not only the dispersion curves of the silicon-air metasurface in Fig.~\ref{fig4}(c) are nearly the same as those of the FCE metasurface in Fig.~\ref{fig2}(b), but also the radiative $Q$ factors in the silicon-air metasurface in Fig.~\ref{fig4}(d) are comparable to the high $Q$ values in the FCE-metasurface in Fig.~\ref{fig2}(c). The radiative $Q$ factors in the silicon-air metasurface are approximately $10^4$ times larger than those in the conventional BDG in Fig.~\ref{fig2}(c). From Fig.~\ref{fig4}, it is reasonable to infer that the high $Q$ values in the FCE metasurfaces are not affected seriously by the small imperfections introduced in the engineered dielectric constant profile.

We note that the proposed concept of eliminating Fourier harmonic components can be extended to zero-contrast grating (ZCG) structures \cite{Magnusson2014,Shokooh-Saremi2014}. Because the ZCG exhibits photonic band gaps and leaky-wave effects due to the spatially modulated thickness profile, the FCE metasurface corresponding to the ZCG can be designed by eliminating the first Fourier harmonic component in the spatially modulated thickness profile. We verified that the radiative $Q$ factors in the FCE metasurface corresponding to the ZCG are noticeably increased owing to the absence of the first Fourier harmonic component in the spatial thickness profile (see the Supplemental Material \cite{Supplemental}). From a practical point of view, manipulation of the spatial thickness profile is more realistic than that of the spatial permittivity profile. Current state-of-the-art nanofabrication technology can be utilized to implement the required spatial thickness profile in the unit cell with submicron size \cite{Staude2017,WTChen2018,Ndao2020}. The Supplemental Material \cite{Supplemental} further presents that the radiative $Q$ factors in the FCE metasurface are not seriously affected by small imperfections introduced in the engineered thickness profile.

In conclusion, we introduced the concept of FCE metasurfaces without the first Fourier harmonic component in spatially modulated lattice parameters, such as the dielectric constant and thickness, and demonstrated that the new types of lattices with 1D and 2D geometries support the continuous quasi-BICs in the vicinity of the second stop bands. The FCE metasurfaces exhibit noticeably increased radiative $Q$ factors, as out-of-plane radiation is attributed to the first Fourier harmonic component of the periodic modulation in lattice parameters. Because the continuous high-$Q$ bound states are robust to small imperfections introduced in the engineered thickness profile and dielectric function, the FCE metasurfaces can be utilized in various applications associated with high-$Q$ resonant Bloch modes. 

\begin{acknowledgments}
This research was supported by grants from the National Research Foundation of Korea funded by the Ministry of Education (Nos. 2020R1I1A1A01073945 and 2018R1D1A1B07050116) and Ministry of Science and ICT (No. 2020R1F1A1050227), along with the Gwangju Institute of Science and Technology Research Institute 2020.
\end{acknowledgments}


\begin{thebibliography}{1}

\bibitem{Kildishev2013} A. V. Kildishev, A. Boltasseva, and V. M. Shalaev, Science {\ bf 339}, 1232009 (2013).
\bibitem{NYu2014} N. Yu and F. Capasso, Nat. Mater. {\bf 13}, 139--150 (2014).
\bibitem{SSun2019} S. Sun, Q. He, J. Hao, S. Xiao, and L. Zhou, Adv. Opt. Photonics {\bf 11}, 38--479 (2019).
\bibitem{Joannopoulos1995} J. D. Joannopoulos, R. D. Meade, and J. N. Winn, \emph{Photonic Crystals: Molding the Flow of Light}, (Princeton University, 1995).
\bibitem{Johnson1999} S. G. Johnson, S. Fan, P. R. Villeneuve, J. D. Joannopoulos, and L. A. Kolodziejski, Phys. Rev. B {\bf 60}, 5751--5758 (1999).
\bibitem{Born2002} M. Born and E. Wolf, \emph{Principles of Optics}, (Cambridge University Press, Cambridge, 2002).
\bibitem{YHKo2018} Y. H. Ko and R. Magnusson, Optica {\bf 5}, 289-294 (2018).
\bibitem{Arbabi2015}  A. Arbabi, Y. Horie, M. Bagheri, and A. Faraon, Nat. Nanotechnol. 10, 937--943 (2015).
\bibitem{AMHWong2018} A. M. H. Wong and G. V. Eleftheriades, Phys. Rev. X {\bf 8}(1), 011036 (2018).
\bibitem{Niraula2015} M. Niraula, J. W. Yoon, and R. Magnusson, Opt. Lett. {\bf 40}, 5062--5065 (2015).
\bibitem{Marinica2008} D. C. Marinica, A. G. Borisov, and S. V. Shabanov, Phys. Rev. Lett. {\bf 100}, 183902 (2008).
\bibitem{Plotnik2011} Y. Plotnik, O. Peleg, F. Dreisow, M. Heinrich, S. Nolte, A. Szameit, and M. Segev, Phys. Rev. Lett. {\bf 107}, 183901 (2011).
\bibitem{Hsu2016} C. W. Hsu, B. Zhen, A. D. Stone, J. D. Joannopoulos, and M. Solja\v{c}i\'{c}, Nat. Rev. Mater. {\bf 1}, 1--13 (2016).
\bibitem{Koshelev2019} K. Koshelev, G. Favraud, A. Bogdanov, Y. Kivshar, and A. Fratalocchi, Nanophotonics {\bf 8}, 725–745 (2019).
\bibitem{SGLee2020-1} S.-G. Lee, S. H. Kim, and C. S. Kee, Nanophotonics {\bf 9}(14), 4373--4380 (2020).
\bibitem{Koshelev2018} K. Koshelev, S. Lepeshov, M. Liu, A. Bogdanov, and Y. Kivshar, Phys. Rev. Lett. {\bf 121}(19), 193903 (2018).
\bibitem{Koshelev2020} K. Koshelev, S. Kruk, E. Melik-Gaykazyan, H.-H. Choi, A. Bogdanov, H.-G. Park, Y. Kivshar, Science {\bf 367}, 288--292 (2020).
\bibitem{BZhen2014} B. Zhen, C. W. Hsu, L. Lu, A. D. Stone, and M. Solja\v{c}i\'{c}, Phys. Rev. Lett. {\bf 113}(25), 257401 (2014).
\bibitem{Doeleman2018} H. M. Doeleman, F. Monticone, W. Hollander, A. Al\`{u}, and A. F. Koenderink, Nat. Photonics {\bf 12}, 397--401 (2018).
\bibitem{Azzam2018} S. I. Azzam, V. M. Shalaev, A. Boltasseva, and A. V. Kildishev, Phys. Rev. Lett. {\bf 121}(25), 253901 (2018).
\bibitem{Yang2014} Y. Yang, C. Peng, Y. Liang, Z. Li, and S. Noda, Phys. Rev. Lett. {\bf 113}(3), 037401 (2014).
\bibitem{JJin2019} J. Jin, X. Yin, L. Ni, M. Solja\v{c}i\'{c}, B. Zhen, and C. Peng, Nature {\bf 574}, 501--504 (2019).
\bibitem{XYin2020} X. Yin, J. Jin, M. Solja\v{c}i\'{c}, C. Peng, and B. Zhen, Nature {\bf 580}, 467–471(2020).
\bibitem{Mermet-Lyaudoz2019} R. Mermet-Lyaudoz, F. Dubois, N. Hoang, E. Drouard, L. Berguiga, C. Seassal, X. Letartre, P. Viktorovitch, and H. S. Nguyen, arXiv:1905.03868 (2019).
\bibitem{YDing2007} Y. Ding and R. Magnusson, Opt. Express {\bf 15}(2), 680--694 (2007).
\bibitem{Yariv1984} A. Yariv and P. Yeh, \emph{Optical Waves in Crystals} (Wiley, New York, 1984).
\bibitem{Inoue2004} K. Inoue and K. Ohtaka, \emph{Photonic Crystals: Physics, Fabrication and Applications} (Springer-Verlag Berlin Heidelberg, 2004.).
\bibitem{Kazarinov1985} R. F. Kazarinov and C. H. Henry, IEEE J. Quant. Electronics {\bf 21}, 144--150 (1985).
\bibitem{SGLee2019-1} S.-G. Lee and R. Magnusson, Phys. Rev. B {\bf 99}(4), 045304 (2019).
\bibitem{Rosenblatt1997} D. Rosenblatt, A. Sharon, and A. A. Friesem, IEEE J. Quant. Electronics {\bf 33}, 2038--2059 (1997).
\bibitem{Supplemental} See Supplemental Material for (1) Semi-analytical dispersion model, (2) Dirac cone dispersions in FCE metasurfaces, (3) Radiative properties of band edge modes in 2D gratings, and (4) FCE metasurfaces corresponding to zero-contrast gratings, which includes Refs.~\cite{Ziolkowski2004,AAlu2007,Liberal2017,Haldane2008,LHWu2015,LLu2014,Minkov2018,BZhen2015}.
\bibitem{Ziolkowski2004} R. W. Ziolkowski, Phys. Rev. E. {\bf 70}(4), 046608 (2004).
\bibitem{AAlu2007} A. Al\`{u}, M. G. Silveirinha, A. Salandrino, and N. Engheta, Phys. Rev. B {\bf 75}, 155410 (2007).
\bibitem{Liberal2017} I. Liberal and N. Engheta, Nat. Photonics {\bf 11}(3), 149--158 (2017).
\bibitem{Haldane2008} F. D. M. Haldane and S. Raghu, Phys. Rev. Lett. {\bf 100}(1), 013904 (2008).
\bibitem{LHWu2015} L. H. Wu and X. Hu, Phys. Rev. Lett. {\bf 114}, 223901 (2015).
\bibitem{LLu2014} L. Lu, J. D. Joannopoulos, and M. Solja\v{c}i\'{c}, Nat. Photonics. {\bf 8}(11), 821--829 (2014).
\bibitem{Minkov2018} M. Minkov, I. A. D. Williamson, M. Xiao, and S. Fan, Phys. Rev. Lett. {\bf 121}, 263901 (2018).
\bibitem{BZhen2015} B. Zhen, C. W. Hsu, Y. Igarashi, L. Lu, I. Kaminer, A. Pick, S.-L. Chua, J. D. Joannopoulos, and M. Solja\v{c}i\'{c}, Nature (London) {\bf 525}, 354--358 (2015).

\bibitem{Srinivasan2002} K. Srinivasan and O. Painter, Opt. Express {\bf 10}, 670--684 (2002).
\bibitem{Peng2012} C. Peng, Y. Liang, K. Sakai, S. Iwahashi, and S. Noda, Phys. Rev. B {\bf 86}(3), 035108 (2012).
\bibitem{Schurig2006} D. Schurig, J. Mock, B. Justice, S. Cummer, J. Pendry, A. Starr, and D. Smith, Science {\bf 314}, 977 (2006).
\bibitem{HFMa2010} H. F. Ma and T. J. Cui, Nat. Commun. {\bf 1}(3), 21 (2010).
\bibitem{Haggans1993} C. W. Haggans, L. Li, and R. K. Kostuk, J. Opt. Soc. Am. A {\bf 10}, 2217--2225 (1993).
\bibitem{Magnusson2014} R. Magnusson, Opt. Lett. {\bf 39}, 4337 (2014).
\bibitem{Shokooh-Saremi2014} M. Shokooh-Saremi and R. Magnusson, Opt. Lett. {\bf 39}, 6958 (2014).
\bibitem{Staude2017} I. Staude and J. Schilling, Nat. Photonics {\bf 11}, 274--284 (2017).
\bibitem{WTChen2018} W. T. Chen, A. Y. Zhu, V. Sanjeev, M. Khorasaninejad, Z. Shi, E. Lee, and F. Capasso, Nat. Nanotechnol. {\bf 13}, 220--226 (2018).
\bibitem{Ndao2020} A. Ndao, L. Hsu, J. Ha, J.-H Park, C. Chang-Hasnain, and B. Kant\'{e} B, Nat. Commun. {\bf 11} 1 (2020).
\end{thebibliography}
\end{document}